 *************************************************************************
 *  All figures for the preprints from Nuclear Physics Group of          *
 *  Drexel University are available through the ftp.                     *
 *  The procedure is : ftp at                                            *
 *        einstein.physics.drexel.edu (129.25.1.120)                     *
 *        anonymous                          (as username)               *
 *        <Return>                           (as password)               *
 *        cd pan_files                       (go to the subdirectory)    *
 *************************************************************************

 For the paper entitled
      Anomalous Crossing Frequency in Odd Proton Nuclei
 The filenames for the four figures are Tafig1.ps, Tafig2.ps, Tafig3.ps
      and Tafig4.ps

\documentstyle[preprint,aps]{revtex}
\begin{document}
\draft
\preprint{\today}
\title{
Anomalous Crossing Frequency
in Odd Proton Nuclei
}
\author{Yang Sun$^{(1)}$,  Shuxian Wen$^{(2)}$
and Da Hsuan Feng$^{(1)}$}
\address {
$^{(1)}$Department of Physics and Atmospheric Sceince, Drexel University \\
Philadelphia, Pennsylvania 19104, USA \\
$^{(2)}$China Institute of Atomic Energy,
P.O.Box 275, Beijing, P.R.China }

\maketitle

\begin{abstract}
A generic explanation for the recently observed anomalous crossing frequencies
in odd proton rare earth nuclei is given. As an example,
the proton ${1\over 2} [541]$
band in $^{175}$Ta is discussed in detail by using
the angular momentum projection theory.
It is shown that the quadrupole pairing interaction is decisive in delaying the
crossing point and the changes in crossing frequency along the isotope chain
are due to the different neutron shell fillings.
\end{abstract}

\newpage

Over twenty years ago, Bes and Broglia introduced the quadrupole pairing force
in their particle-vibration-coupling model~\cite{BB.71}. It was shown later
that such a force will tend to attenuate the Coriolis interaction in an odd
mass system~\cite{Hama.74,HK.75} and shift the band crossing point in an
even-even system~\cite{WF.78,HI.80}. Curiously the effect of this force on
the odd mass band crossing did not receive much attention.  The lack of which,
even to this date, may be due partly to the fact that there is no relevant data
appropriate for this investigation. In the past few years, a large amount of
high spin data were obtained for odd proton isotopes (Lu, Ta, Re and Ir) and
they systematically showed an anomalous shift in the crossing frequency (see
\cite{Ta169,Ir} and references therein). Quite recently, an especially large
crossing frequency for the proton band labeled by the Nilsson quantum numbers
${1\over 2} [541]$ in $^{175}$Ta was observed at Beijing's Institute of Atomic
Energy Tandem Laboratory \cite{Wen.91}. In addition, it was also observed that
going from light to heavy Ta isotopes, there is a drastic change in the
crossing frequency. With this set of new data, the situation is riped to carry
out a study of the above mentioned problem. In this letter we shall discuss
the physics mainly by using $^{175}$Ta as an example. However, it is worth
emphasizing that the discussions could be equally applied to other odd proton
nuclei as well.

It is well known that in rare earth nuclei, the crossing of bands (i.e. two
bands with rather different rotational frequencies cross at a certain angular
momentum) can be interpreted as the alignment of a pair of $i_{13\over 2}$
quasineutrons along the rotational axis. For $^{175}$Ta the crossing frequency
was found to be 0.375 MeV/$\hbar$. This significant increase in value from its
even-even neighbors (0.292 MeV/$\hbar$ for $^{174}$Hf and 0.291 MeV/$\hbar$ for
$^{176}$W) indicates that there must be a physical mechanism to delay the
crossing of the bands. Indeed the problem is made more acute when Wen et al.
\cite{Wen.91} found that within the framework of the semi-classical cranking
shell model, one needs to use an unreasonably large quadrupole deformation
$\epsilon _2$ of 0.3 to account for this observation.
However, there is a lingering question as to whether the cranking
theory is useful to discuss the phenomena of band crossing since in the
cranking approximation it admixes bands at a given rotational frequency (not at
a given angular momentum), thus giving rise to large uncertainty in the band
crossing region~\cite{Hama.76}.

Clearly, the study of the problem requires a model which can treat the band
crossings quantum mechanically. Shell model configuration mixing calculations
could in principle solve this problem, but in practice is unfeasible for heavy
systems. The angular momentum projection theory established in the late
seventies~\cite{HI.79,HI.80} has been demonstrated to be a powerful model to
quantitatively account for many high spin phenomena~\cite{HS.91a,HS.91b,HS.92}.
The development of an efficient algorithm~\cite{HI.79} renders it possible, in
a unified manner, to systematically investigate the even-even, odd-odd and odd
mass heavy systems. Hence the model is particularly suitable for the present
study. Since the model has already been extensively discussed
\cite{HI.80,HS.91a}, only the salient features will be given below. We should
add that a similar approach to the projection technique was also developed by
the T\"ubingen group~\cite{SG.87}.

The ansatz for the angular momentum projected wave function is given by
\begin{equation}
| I M > ~=~
\sum_{\kappa} f_{\kappa} \hat P^I_{MK_\kappa}
| \varphi_{\kappa} > ,
\label{ansatz}
\end{equation}
where $\kappa$ labels the basis states. Acting on an intrinsic state $|
\varphi_{\kappa} >$, the projection operator
$\hat P^I_{MK}$ \cite{RS.80} generates states
of good angular momentum, thus restoring the necessary rotational symmetry
which was violated in the deformed mean field. The advantage of the present
approach is that the crossing and mixing of bands at a given angular momentum
are treated fully quantum mechanically. This turns out to be crucial to treat
the present problem.

In the present work, we assume that the intrinsic states have axial symmetry.
Thus, the basis states $| \varphi_{\kappa} > $ must have $K$ as a good quantum
number.  Since the nuclei in question have only very weak $\gamma$ deformation,
this restriction shall not prevent us to investigate the physics at hand. The
basis states $| \varphi_{\kappa} > $ are spanned by the set
\begin{eqnarray}
\left\{
\;\; \alpha^\dagger_{p_l} |\phi >,
\;\;\; \alpha^\dagger_{n_i} \alpha^\dagger_{n_j} \alpha^\dagger_{p_l}
|\phi > \;\; \right\}
 \nonumber \\
\left\{\;\; |\phi >, \;\;\; \alpha^\dagger_{n_i} \alpha^\dagger_{n_j} |\phi >,
\;\;\; \alpha^\dagger_{p_k} \alpha^\dagger_{p_l} |\phi >,
\;\;\; \alpha^\dagger_{n_i} \alpha^\dagger_{n_j} \alpha^\dagger_{p_k}
\alpha^\dagger_{p_l} |\phi >\;\; \right\}
\label{baset}
\end{eqnarray}
for odd proton and even-even nuclei, respectively. The quasiparticle
vacuum is $|\phi >$ and $\{\alpha_m, \alpha^\dagger_m \}$ are the quasiparticle
annihilation and creation operators for this vacuum; the index $n_i$ (~$p_i$~)
runs over selected neutron (proton) quasiparticle states and $\kappa$ in
eq.~(\ref{ansatz}) runs over the configurations of eq.~(\ref{baset}). The
vacuum $|\phi >$ is obtained by diagonalizing a deformed Nilsson hamiltonian
\cite{An.78} followed by a BCS calculation. In the calculation, we have used
three major shells: i.e. N = 4, 5 and 6 (N = 3, 4 and 5) for neutrons (protons)
as the configuration space. For the odd system, the BCS blocking effect
associated with the last unpaired proton is approximately taken into account by
allowing all the odd number of protons to participate without blocking any
individual level. Thus the vacuum in this case is an average over the two
neghboring even-even nuclei. The size of basis states, which includes the
most important configurations, is determined by using energy windows of
1.5 MeV, 2.5 MeV, 4 MeV and 5 MeV for the 1-, 2-, 3- and 4-qp states,
respectively.

In this work, we have used the following hamiltonian
\cite{HI.80}
\begin{equation}
\hat H = \hat H_0 - {1 \over 2} \chi \sum_\mu \hat Q^\dagger_\mu
\hat Q^{}_\mu - G_M \hat P^\dagger \hat P - G_Q \sum_\mu \hat
P^\dagger_\mu\hat P^{}_\mu ,
\label{hamham}
\end{equation}
where $\hat H_0$ is the spherical single-particle shell model hamiltonian.
The second term is the
quadrupole-quadrupole interaction and the last two terms are the monopole and
quadrupole pairing interactions respectively. The interaction strengths
are determined as follows: the quadrupole interaction strength $\chi$ is
adjusted so that the known quadrupole deformation $\epsilon _2$ from the
Hartree-Fock-Bogoliubov self-consistent procedure \cite{Lamm.69} is obtained.
For example, for $^{175}$Ta it is 0.26; the monopole pairing strength
$G_M$ is adjusted to the known energy gap
\begin{equation}
G_M =  \left[ 20.12\mp 13.13 \frac{N-Z}{A}\right] \cdot A^{-1} ,
\label{GMONO}
\end{equation}
where the minus (plus) sign is for neutrons (protons). The quadrupole pairing
strength $G_Q$ is assumed to be proportional to $G_M$ and the proportional
constant is typically $C \approx$ 0.20. This is the only adjustable parameter
in the present model.

The weights $f_{\kappa}$ in eq.~(\ref{ansatz})
are determined by diagonalizaing the
hamiltonian $\hat H$ in the basis given by of eq.~(\ref{baset}).
This will lead to
the eigenvalue equation (for a given spin $I$)
\begin{equation}
\sum_{\kappa '} ( H_{\kappa\kappa '} - E N_{\kappa\kappa '} )
f_{\kappa '} ~=~ 0 ,
\label{maha}
\end{equation}
with the hamiltonian and norm overlaps given by
\begin{eqnarray}
H_{\kappa \kappa'} &~=~& < \varphi_\kappa |\hat H \hat
P^I_{K_\kappa {K'}_{\kappa'}}|\varphi_{\kappa'} >,
 \nonumber \\
N_{\kappa\kappa'} &~=~& < \varphi_\kappa | \hat
 P^I_{K_\kappa {K'}_{\kappa'}} | \varphi_{\kappa'} > .
\label{norm}
\end{eqnarray}

Projection of good angular momentum onto each intrinsic state generates the
rotational band associated with this intrinsic configuration
$| \varphi_{\kappa}>$. For example, $\hat P^I_{MK} \alpha^\dagger_{p_l} |\phi>$
will produce a one-quasiproton band. The energies of each band are given by
the diagonal elements of eq.(\ref{norm})
\begin{equation}
E_\kappa (I) ~=~ \frac{ <\varphi_\kappa | \hat H \hat P^I_{KK} |
\varphi_\kappa >}{ < \varphi_\kappa | \hat P^I_{KK} |
\varphi_\kappa > } ~=~ \frac{ H_{\kappa\kappa}} {N_{\kappa\kappa} } .
\label{bandiag}
\end{equation}
A diagram in which $E_\kappa (I)$ of various bands are plotted against
the spin $I$ will be referred to \cite{HS.91a} as a band diagram. It will
reveal information for understanding the character of the observed band
crossings. The results obtained from diagonalizing the hamiltonian of
eq.(\ref{hamham}) can be compared
with the experiments.

In fig.1, the band diagram for negative parity bands of $^{175}$Ta is
presented.  Although there are many bands in the calculation,
only four most interesting bands will be plotted in order
to illustrate the main features.
The rotational frequency of each band, $\omega (I) = {d E(I)\over {d I}}$,
is naturally described by the slope of the curve. Its inverted value gives
the moment of inertia. In fig.1, one can see that at a certain angular
momentum,
different configurations have different slopes.
The band which is labeled by the Nilsson quantum
numbers $9\over 2$[514] shows the usual smooth behavior as a function of
increasing angular momentum. From the figure, we see that it roughly crosses
with the $1\over 2$[541] band at spin ${21\over 2} \hbar$ and continues
upward monotonously. At about spin ${31\over 2} \hbar$, it enters into a
region where several bands converge and interact with each other. At this
point, the experimental assignments~\cite{Wen.91} of the levels can no longer
be made in a clear cut manner.
We predict that the $9\over 2$[514] band
will cross the 3-qp band at spin ${37\over 2} \hbar$.
We anticipate that one should
be able to observe this band crossing if
the present data \cite{Wen.91} for the $9\over 2$[514]
band is extended to higher spins.

For the $1\over 2$[541] band, the zig-zag
behavior indicates a strong signature splitting in energy.
In fact, only the favored branch
with signature $1\over 2$ has been experimentally observed.
This one-quasiproton band clearly crosses the 3-qp band
at spin ${45\over 2} \hbar$, thus producing the observed anomaly in
the spectra. After this crossing, the structure of the Yrast band should mainly
be 3-qp in nature.
It should be pointed out that without any additional assumption,
our calculation clearly indicates that
the $1\over 2$[541] band
crosses the 3-qp band at a much later stage
(spin ${45\over 2} \hbar$) than the
$9\over 2$[514] band (spin ${37\over 2} \hbar$).

In fig.2a we compare the results after diagonalization
with the data~\cite{Wen.91} for
the proton ${1\over 2}$[541] band. This can be succinctly presented by
plotting the rotational frequency as a function of the angular momentum. As can
be seen from this figure, the theory agrees well with the data. In particular,
the rotational alignment at spin ${45\over 2} \hbar$ is reproduced. It is
important to notice that the quadrupole pairing force in the hamiltonian of
eq.~(\ref{hamham}) is crucial here. Its influence on the results
is demonstrated in
fig.2b. Indeed, by increasing the $C$ from 0.16 to
0.24, a clear delay of the alignment process is obtained. If this force is not
included in the hamiltonian (i.e. $C$ =  0), then the alignment could occur as
early as spin ${33\over 2} \hbar$. The physical reason behind the delay is as
follows: If a zero angular momentum pair is broken in the absence of
quadrupole pairing, then there exists no additional force to resist the
alignment process beyond that point ~\cite{WF.78}. In other words, the
quadrupole pairing interaction prevents the alignment from occuring too soon.
Suffice to mention here that the positive parity bands for $^{175}$Ta have also
been computed and they too agree well with the data for all the known
bands~\cite{Wen.91}. Detail discussions will be published elsewhere.

It is natural to inquire whether the drastic changes in the crossing frequency
along the Ta isotope chain requier the readjustment of the quadrupole pairing
force along the chain since this could be a judgement of the
present model. It is therefore gratifying that without changing any of
the parameters (in particular, keeping $C$ = 0.24), we can reproduce the
observed crossing frequencies of the isotopes $^{167, 169, 171, 173}$Ta, as
shown in fig.3.  We should mention that for the two lighter
isotopes $^{167}$Ta and $^{169}$Ta, although the crossing positions are
correctly predicted, there is too much alignment. In addition, for $^{169}$Ta
and $^{171}$Ta, the deviation between theory and data at high spins from
${57\over 2} \hbar$ onwards can be attributed to the missing configurations of
the 5-qp states in the computer code. One believes that
crossings between 3-qp and 5-qp bands can occur here~\cite{Ta169}.

The different crossing frequencies for various isotopes are
simply due to the neutron shell fillings.  We know that the proton
Fermi level remains nearly identical for all the isotopes. Therefore, the
energy and the character of the projected one-quasiproton state
should remain unchanged. However, because of the differences of the neutron
Fermi levels, the energies and the configurations of the
additional neutron pair in the projected
3-qp states can and will change with neutron
number. This can alter the behavior along the isotopic chain of the 3-qp bands
in the energy versus angular momentum band diagram. In fact, they can differ in
energies (the bands lie higher or lower) and/or
rotational frequencies (the slope of bands are steeper or flatter).
Consequently the crossing positions between the proton $1\over 2$[541]
and the 3-qp bands and their interactions are quite different. Thus in
hindsight it is not surprising that there are drastic changes in the crossing
frequencies along the chain.

As we have mentioned at the beginning, the delay of the crossing frequency
is measured by comparing an odd mass nucleus with its even-even neighbors.
Hence a unified treatment will demand us to examine the even-even neighboring
nuclei with the same theory as well. In fig.4, we present our results for
$^{174}$Hf and $^{176}$W. By varying the quadrupole pairing strength, one
observes the effect of shifting the crossing points. However, the effect is
clearly not as significant as in the odd mass Ta case.
It seems that the quadrupole pairing force
is much less sensitive to the even system. Hence the theory without
any re-adjustment in the interacting strength, can consistently describe band
crossings for various systems.
Furthermore, we notice that
the effect is less pronounced at very low spins than it at higher spins,
as one can see from fig. 2b. Further investigations of this
point is clearly necessary.

In conclusion, we have studied the anomalous crossing frequency observed
in $1\over 2$[541] band in the Ta isotopes. It has been shown that by using
the projection theory and including the quadrupole pairing interaction
in the hamiltonian, one can reproduce the essential physics here. In particular
the crossing frequencies of all five Ta isotopes, especially the delay in
alignment in $^{175}$Ta can be described in a unified manner and is a natural
consequence of certain neutron shell filling. The influence of the quadrupole
pairing force on the isotons $^{174}$Hf and $^{176}$W has also been discussed.
This is an example of how one could amalgamate in a unified manner
in one system the two seemingly unrelatated effects of the quadrupole pairing
force,
namely attenuating the Coriolis force
and shifting the band crossings found
in early studies. Furthermore,
from this study it suggests that the high spin region may just be the
sensitive window to determine accurately one of the
effective interactions, i.e. the
quadrupole pairing interaction.

\acknowledgments
This work is partially supported by the United States National Science
Foundation.

\baselineskip = 14pt
\bibliographystyle{unsrt}

\begin{figure}
\caption{Band diagram of $^{175}$Ta. Two proton 1-qp bands $9\over 2$[514] and
$1\over 2$[541] and the corresponding 3-qp bands are plotted.
In the calculation, the quadrupole pairing strength $C$ = 0.24 is used.}
\label{figure.1}
\end{figure}
\begin{figure}
\caption{Rotational frequency versus angular momentum plot for
the proton $1\over 2$[541] band in $^{175}$Ta. a) Top:
Comparison of the calculation
with data~\protect\cite{Wen.91}. b) Bottom: Influence of the
quadrupole pairing force on the crossing frequency.
}
\label{figure.2}
\end{figure}
\begin{figure}
\caption{Rotational frequency versus angular momentum plot for
the proton $1\over 2$[541] band for four Ta isotopes.
Data are taken from: $^{167}$Ta~\protect
\cite{Ta167}, $^{169}$Ta~\protect\cite{Ta169},
$^{171}$Ta~\protect\cite{Ta171} and $^{173}$Ta
\protect\cite{Ta171}.
In the calculation, the quadrupole pairing strength $C$ = 0.24 is used.
}
\label{figure.3}
\end{figure}
\begin{figure}
\caption{Rotational frequency versus angular momentum plot for
the yrast band of the two isotons
$^{174}$Hf and $^{176}$W. Data are taken from:
$^{174}$Hf~\protect\cite{Hf174} and  $^{176}$W~\protect\cite{W176}.
The influence of the
quadrupole pairing force on the crossing frequency are shown.
}
\label{figure.4}
\end{figure}
\end{document}